\documentclass[aps,preprint,noshowpacs,nofootinbib,superscriptaddress,groupedaddress,floatfix,superscriptaddress,preprintnumbers,10pt,twocolumn]{revtex4}  
\usepackage{graphicx} 
\usepackage{dcolumn}   
\usepackage{bm}        
\usepackage{amssymb}   
\usepackage{hyperref}
\usepackage{multirow}
\usepackage{feynmf}
\usepackage{amsmath,amssymb}
\usepackage{graphicx,bm}
\usepackage{slashed}
\usepackage{natbib}
\usepackage{subfigure}
\usepackage{epstopdf}
\usepackage{ulem} 
\usepackage[usenames]{color}

\usepackage{float}

\renewcommand\sout{\bgroup \color{red} \ULdepth=-.5ex \ULset}

\newcommand{\be}{\begin{equation}}
\newcommand{\ee}{\end{equation}}
\newcommand{\ba}{\begin{eqnarray}}
\newcommand{\ea}{\end{eqnarray}}

\begin{document}
\begin{flushright}
\end{flushright}

\preprint{INT-PUB-23-003}

\title{Quarkyonic Mean Field Theory}

\author{Dyana C. Duarte}
\affiliation{Departamento de F\'isica, Universidade Federal de Santa Maria, Santa Maria, RS 97105-900, Brazil}
\affiliation{Departamento de F\'{i}sica, Instituto Tecnol\'ogico de Aeron\'autica, S\~ao Jos\'e dos Campos, SP, 12228-900, Brazil}

\author{Saul Hernandez-Ortiz}
\affiliation{Department of Physics and Astronomy, Iowa State University, Ames, IA 50010}
\affiliation{Instituto de F\'isica y Matem\'aticas, Universidad Michoacana de San Nicol\'as de Hidalgo, Edificio C-3, Ciudad Universitaria, Francisco J. M\'ujica S/N Col. Fel\'icitas del R\'io, 58040 Morelia, Michoac\'an, M\'exico.}
\author{Kie Sang Jeong}
\author{Larry D. McLerran}
\email[]{mclerran@me.com}
\affiliation{Institute for Nuclear Theory, University of Washington, Box 351550, Seattle, WA, 98195, USA}

\date{\today}

\begin{abstract}
We discuss mean field theory of Quarkyonic matter at zero temperature.  We treat the nucleons with contact interactions in mean field approximation, discussing both vector and scalar mean field interactions. We treat the quarks without mean field vector interactions, but allow mass terms to be generated consistent from a scalar mean field consistent with the additive quark model for quark masses.  Quarkyonic matter is composed of a shell of nucleons that under-occupy the total available phase space associated with the underlying quark degrees of freedom.  The fully occupied Fermi sphere beneath this shell of nucleons at high densities is thought of as quarks, but when this fully occupied distribution of states first appears, although the phase space is filled, the matter is at low density.  For the transition between this low density and high density saturated matter, we advocate a dual description of the fully filled Fermi sea in terms of hadrons, and make a phenomenological hypothesis for the equation of state of this matter.  We then proceed to an example where the mean field interactions are all vector and only associated with the nucleons, ignoring the effects of mass change associated with the scalar interactions.  Except for the effects of Pauli blocking, the nucleons and quarks do not interact.  To get a reasonable transition to Quarkyonic matter the interaction of the quarks among themselves are assumed to be non-perturbative, and a simple phenomenological relation between quark Fermi energy and density is introduced.

\end{abstract}


\maketitle

\section{Introduction}

In a recent paper, we argued that Quar\-ky\-o\-nic Matter\cite{McLerran:2007qj} might be described by a field theory with nu\-cleon, ghost and quark degrees of freedom\cite{Duarte:2021tsx}.  The ghosts are introduced to avoid double counting of states where the quarks inside of nucleons might occupy the same physical states as those associated with quarks.  Such a picture has been clearly elaborated in the work of Kojo~\cite{Kojo:2021ugu,Kojo:2021hqh}, and was the starting point of various pre\-vi\-ous studies~\cite{McLerran:2018hbz,Jeong:2019lhv,Sen:2020peq,Cao:2020byn,Kovensky:2020xif}.  

In the field theory with ghosts, there are three chemical potentials that need to be determined, the nucleon chemical potential that are $\mu_N$, the ghost chemical potential, $\mu_G$ and the quark chemical potential $\mu_Q$. The ghost chemical potential is determined in terms of the quarks to avoid double counting of states.  The quark chemical potential is determined by extremizing the pressure of the system, or equivalently at zero temperature, minimizing the energy per nucleon at fixed total baryon number.  Finally, the nucleon chemical potential is determined by  the total baryon number of the system.

It is the purpose of this paper to understand how all of this might work in detail in a simple model.  We will consider a mean field theory of nucleon interactions interacting with a vector potential.  We show how a scalar mean field might modify these considerations, but we explicitly consider only the case of vector mean field interactions. The quarks are introduced by low energy QCD effective model. In our simple model, they are allowed to interact among themselves but not with the nucleons.    Their interactions are strong, and modify the free quark relation between density and chemical potential.  We will introduce a simple phenomenological parameterization of this relationship. The ghost fields enforce the constraint that nucleons and quarks do not occupy the same phase space.  At high density, the quarks are an almost free gas of quarks, but at low density when they first appear, the filled Fermi sea of quarks may be thought of as a gas of nucleons and their excited states that completely fill the quark energy levels.  

We then turn to the issue of computing the properties of Quarkyonic matter in the limit of a large number of colors $N_c$~\cite{tHooft:1973alw,Witten:1979kh}. We show that if interactions of quarks are ignored, and if the constituent quark masses are $M_Q = M_N/N_c$  and the nucleons are treated in mean field, then Quarkyonic Matter does not form.  There are only two phases:  one of nucleonic matter at low density and the other of quarks at high density.  

We then argue that the region where all the quark states are filled may be thought of as matter state composed of the nucleon and nucleon resonances where the interdistance between the baryons is short so that the quasi-baryon properties can be approximated by the constituent quark dynamics (quark-hadron duality). In this configuration, all the available quark state looks fully occupied since the independent constituent quark wave functions from the nucleon ground state and higher resonances interfere with each other and double occupancy of same quantum number is prohibited by Pauli principle.   We then show that in this circumstance, if this exotic state is treated as a non-interacting  gas of constituent quarks, then there is a first order phase transition to Quarkyonic Matter.  

When the compressed state understood as the filled quark sea first appears, the typical momentum scale is small and interactions cannot be ignored. We construct a simple model of the properties of this low density matter in a density expansion around zero density.  We argue that it is possible that the effect of such interactions may convert the transition between nucleonic matter into Quarkyonic matter as a continuous transition.  This continuous transition is needed to phenomenologically explain the rapid rise of the sound velocity extracted from equations of state of nuclear matter appropriate for neutron stars~\cite{Drischler:2021bup,Tews:2018kmu,Steiner:2017vmg,Bedaque:2014sqa}.

We should emphasize at the outset, that our goal is not to provide a phenomenologically viable model of Quarkyonic matter.
This paper is simply an exploration of how Quarkyonic matter might appear in a field theory with both nucleon and quark degrees of freedom present, but not allowing simultaneous multiple occupation of phase space of the quarks within nucleon with those of the filled Fermi sea of quarks.  It is a first very small step towards constructing viable theories at finite density and temperatures, that properly include the effect of interactions, and the low occupation number nucleon states at the Fermi surface.

\section{Review of Mean Field Theory for Applications to Quarkyonic Matter}

Before proceeding to a mean field description of Quarkyonic Matter, we review the basic ingredients of mean field theory that we will use in the following. We will take all of the nucleon interactions to be given by contact interactions.  First consider a single species of baryons at finite density.  The theory with a scalar in\-te\-rac\-tion is
\begin{equation}
 S = \int ~d^4x \left\{  \overline \psi \left({1 \over i} \slashed \partial - \gamma^0\mu^* + M - g_s \sigma \right) \psi  + {M_s^2 \over 2} \sigma^2\right\}\,. 
\end{equation}
In the entire paper, $\mu^*$ denotes the bare chemical potentials, without any kind of interactions. The inclusion of the scalar field generates a nucleon-nucleon contact interaction.  Integrating out the scalar field gives
\begin{equation}
 S = \int ~d^4x \left\{  \overline \psi \left({1 \over i} \slashed \partial - \gamma^0\mu^* +M \right) \psi  - {g_s^2 \over {2 M_s^2}}  (\overline \psi \psi )^2\right\} \,.
\end{equation}
In mean field approximation, we replace
\begin{equation}
\overline \psi \psi \rightarrow \overline \psi \psi + n_{s}\,,
\end{equation}
where 
\begin{equation}
n_{s} = \langle\overline \psi \psi\rangle\,.
\end{equation}
The action becomes
\begin{eqnarray}
  S &=& \int ~d^4x \Biggl\{  \overline \psi \left({1 \over i} \slashed \partial - \gamma^0\mu^* +M_{\text{eff}} \right) \psi \nonumber\\
  &&- {g_s^2 \over {2 M_s^2}}  \left[(\overline \psi \psi )^2 + \langle\overline \psi \psi \rangle^2 \right] \Biggl\} \,.
\end{eqnarray}

The mean field approximation consists of including the explicit term that involves the contribution of 
$\langle\overline \psi \psi\rangle^2$ and  the contribution of the ideal gas term with the $\langle\overline \psi \psi\rangle$ computed in mean field approximation. For the energy density, we obtain
\begin{equation}
  \epsilon = \epsilon_{\text{kin}} - {g_s^2 \over {2 M_s^2}} n_s^2\,.
\end{equation}
The kinetic energy is
\begin{equation}
\epsilon_{\text{kin}}= \int {{d^3p} \over {(2\pi)^3}} ~n(p)~\sqrt{p^2+M^2_{\text{eff}}}\,,
\end{equation}
and
\begin{equation}
   n_s = \int~{{d^3p} \over {(2\pi)^3}}~ n(p)~{M_{\text{eff}} \over \sqrt{p^2 + M_{\text{eff}}^2} }\,.
\end{equation}

For the vector field treatment, the chemical potential is the maximum energy difference beyond the ground state energy.  The vector potential shifts the overall zero of energy by the interactions energy with the zeroth component of the vector field $g_v V^0$. The vector field shifts the ground state energy of the fermion by a constant in mean field approximation. An overall shift by a constant does not affect the one loop contribution to the action.  The only effect is to include the interaction energy in the mean field energy functional.  So the result is
\begin{equation}
 \epsilon = \epsilon_{\text{kin}} + {g_v^2 \over {2M_v^2}} n_v^2- {g_s^2 \over {2 M_s^2}} n_s^2\,,
\end{equation}
where
\begin{equation}
  n_v = \int~{{d^3p} \over {(2\pi)^3}} (n_{\text{particle}} - n_{\text{antiparticle}})\,.
\end{equation}
At zero temperature, which is the case that we consider through the rest of this work
 \begin{equation}
  n_v = \int~{{d^3p} \over {(2\pi)^3}}\,.
\end{equation} 
For purposes of $N_c$ counting, one should note that  $g_s^2$ and $g_v^2$ are both of order $N_c$.

When we generalize this discussion to Quarkyonic matter, we add in ghost fields.  The ghost fields are bosonic spinor fields. The Lagrangian for the nucleon, including the effect of ghost is  
\begin{eqnarray}
 S & = & \int ~d^4x \Biggl\{  \overline \psi \left({1 \over i} \slashed \partial - g_v\slashed V - \gamma^0\mu_N^* + M - g_s \sigma \right) \psi \nonumber\\
 & & + \overline G \left({1 \over i} \slashed \partial - g_v\slashed V - \gamma^0\mu_G^* +M -g_s \sigma \right) G \nonumber   \\
 &  &  + {M_s^2 \over 2} \sigma^2 + {M_V^2 \over 2}  A^2 \Biggl\}\,.
\end{eqnarray}
In mean field theory, the energy function for nucleons is modified by the ghost by $n_v \rightarrow n^N_v - n^G_v$ and
$n_s \rightarrow n_s^N - n_s^G$.

For Quarkyonic matter, we need the contributions of the nucleons at chemical potential $\mu_B$, ghosts at chemical potentials $\mu_G$. The quarks at chemical potential $\mu_Q$ will be introduced via QCD effective model. In what follow, we will ignore the QCD interactions of quarks with ghost and nucleons.  They know each other only through the Fermi exclusion principle, and this  will be implemented by requiring the ghost potential is
\begin{equation}
  \mu_G =  N_c \mu_Q + g_vV^0\,,
\end{equation}
where $V^0 = (g_v/M_v^2)n_v$ is the background vector potential.  This means that ghost states block the states of nucleons where their kinetic momentum is $k_G = N_c k_Q$, at least in the limit where the constituent quark mass is $M_Q = M_N/N_c$.
If the scalar mean field is included the nucleon develops an effective mass $M_{\text{eff}} = M_N - g_s\sigma$ that is a function of the baryon density. In this case, the scalar condensate $\sigma$ would be obtained by extremizing the pressure with respect to the nucleon effective mass to obtain
\begin{equation}
 \sigma = \frac{g_s}{m_s^2}n_s\;.
\end{equation}

In what follows, we will consider several cases.  In all we consider, we ignore the effect of a scalar mean field.  A scalar field could be implemented in more detailed models. We first consider the case where quarks are free and have $M_Q = M_N/N_c$.  We then  add  $\Delta M_Q$ to the quark mass, and finally we will develop a phenomenological model of the quark interactions. 
The reasons for these increasing levels of complication arise because we want to eventually obtain a model where there is a continuous transition between nucleonic matter and Quarkyonic matter.  The motivation for these generalizations and their consequences are the subject of the following sections.

\section{An Ideal Gas of Quarks}

 At very high density, the picture of Quarkyonic Matter in momentum space is that of a thin shell of nucleonic matter which surrounds a filled Fermi sea of quarks.  At very high density, the typical momentum of the Fermi sphere of quarks is large compared to the QCD scale, and one can assume that the gas is a quasi-free gas of quarks, with interactions controlled by a coupling which is small at the scale of interest.
 
 When Quarkyonic matter first appears however, we must be careful in such considerations.  As we approach high density in a gas of nucleons, excited nucleon states begin to occur.  They of course fill states with minimal Fermi momenta, in order to minimize the energy as they first appear. Up to the intermediate densities, the system may have some independent Fermi sea of baryons. As yet higher energy density is approached, the ground state nucleons and the resonances get closer and the cons\-ti\-tuent quark wave functions start to interfere with each other.  These excited nucleon states should not be viewed like thermal excitations, but states with different wave functions of the quark degrees of freedom so that they can fill up all the available vacant quark energy levels in such a compressed matter configuration. This will continue until a density is achieved where all of the quark states at lower momentum in the Fermi sphere are filled.  In such a picture, there is indeed a Fermi surface of nucleon states.  The Fermi sphere, eventually corresponding to all the quark states appearing, is a Fermi sphere surrounded by nucleons.  If instead of thinking about nucleon and excited nucleon degrees of freedom, we think of the states occupied by excited nucleon states as simply in terms of quark degrees of freedom, we recover the Quarkyonic description.  The excited nucleon states are dual to the quark description, and thinking in terms of excited nucleon states is probably most convenient when the nontrivial states first appear, and the typical momentum scale is not large compared to the QCD scale.  The picture we advocate here is essentially that from Refs.~\cite{Kojo:2021ugu, Kojo:2021hqh, Park:2019bsz, Park:2021hqb}.
 
 We can imagine various approximate descriptions of this Fermi sphere of filled quark states.  First, in the analysis below, we consider the filled Fermi sphere to be that of constituent quarks, with non-interacting quarks and the quark constituent mass to be $M_Q = M_N/N_c$.  The nucleons self interact with a vector mean field.  
 
 We will set the coupling of the scalar field for quarks to be $1/N_c$ times that for the scalar field for the nucleons and ghosts, so that the additive quark parton model is preserved at any density, $M_N = N_c M_Q$.  We ignore the vector interaction for quarks, as we expect that the scale dependent self interaction of nucleons will be replaced by interactions terms of quark with gluons.   A proper treatment of the quarks for the vector interaction must inevitably involve a proper treatment of quarks and gluons.   The energy density of the system of quarks, gluons  and ghosts  is:

\begin{eqnarray}
 \epsilon & = & \epsilon_{\text{kin}}^N(\mu_B) - \left(1 - {1 \over N_c^3}\right) \epsilon_{\text{kin}}^N(N_c\mu_Q)  \nonumber \\
 & & + {g_v^2 \over {2M_v^2}}(n_v(\mu_B) - n_v(N_c\mu_Q))^2\nonumber\\
 & & -{g_s^2 \over {2 M_s^2}}\left(n_s(\mu_B) - \left(1 - {1 \over N_c^3}\right)n_s(N_c\mu_Q)\right)^2.\nonumber\\
 \label{EnDens}
\end{eqnarray}
The nucleon energy density is $\epsilon^N$ and we used the scaling relation for the masses of quarks to relate the quark energy density at chemical potential $\mu_Q$ to that of the ghosts at chemical potential $N_c \mu_Q$~\cite{Duarte:2021tsx}.
There is a constraint on the total baryon number:
\begin{equation}
n_B = n_v(\mu_B) - \left(1 - {1 \over N_c^3}\right)n_v(N_c\mu_Q)\,.
\end{equation}

Since the matter we consider in the filled Fermi sphere can be thought of as the compressed state of the ground state nucleon and resonances, it is reasonable to assume that a description in terms of constituent quarks has constituent quarks masses $M_Q= M_N/N_c + \Delta M$.  In this case one should shift the ghost chemical potential by the threshold mass 
 \begin{equation}
  \mu_G = N_c(\mu_Q - \Delta M) + g_v V^0
 \end{equation}
so that we begin having ghost states when the quark states appear~\cite{Duarte:2021tsx}.
\begin{figure*}[t!]
\begin{center}
 \subfigure{\includegraphics[scale=0.35]{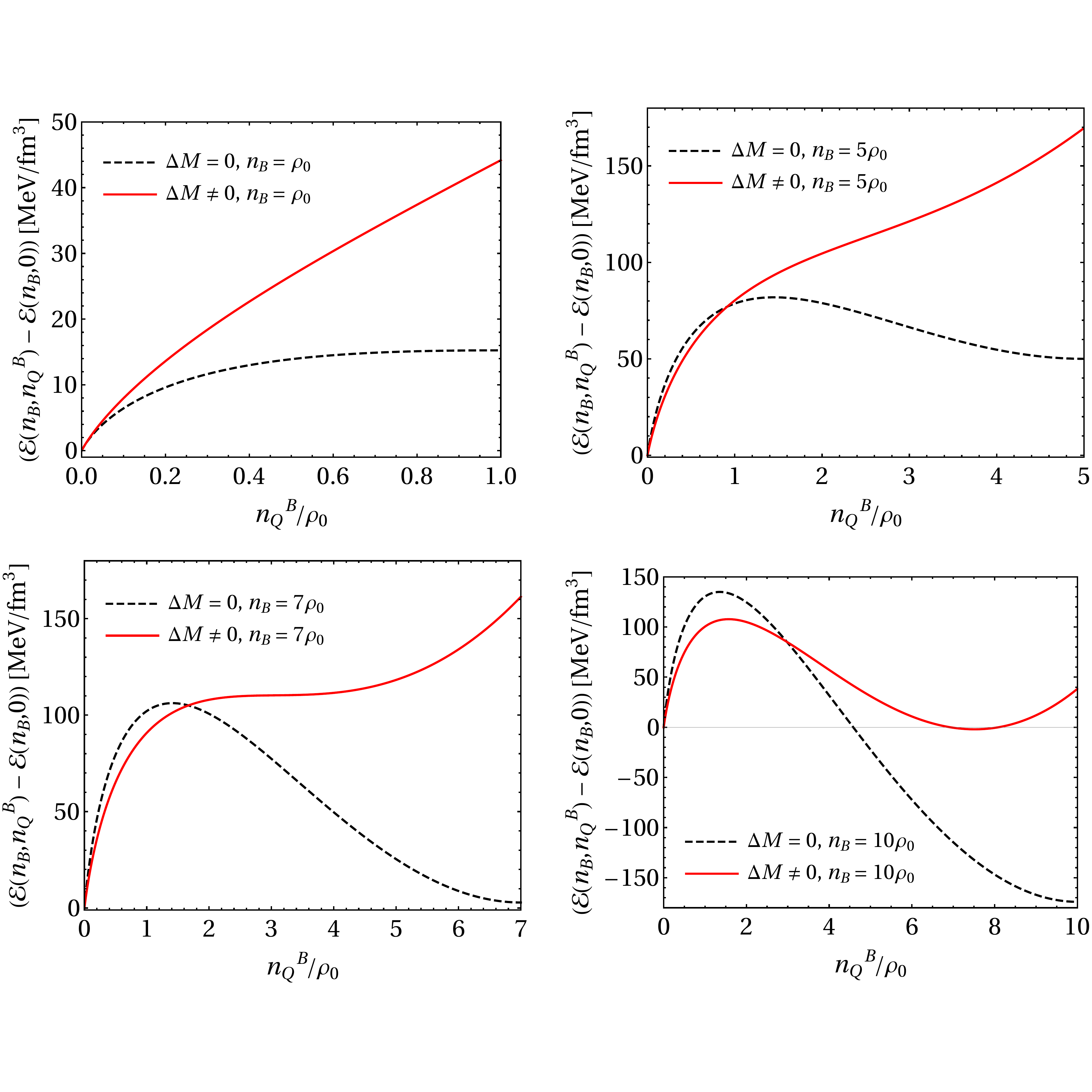} }
\caption{Energy density as a function of quark density $n_Q^B = n_Q/N_c$ including only vector interactions, for the free gas of quarks comparing $M_Q= M_N/N_c$ (dashed) and $M_Q= M_N/N_c + \Delta M$ (solid), for different values of total baryon density $n_B$. Here we have used $\Delta M = 70$ MeV.}
\label{fig1}
\end{center}
\end{figure*} 
In Fig.~\ref{fig1} we plot the energy density as a function of quark density for both $M_Q= M_N/N_c$ and $M_Q= M_N/N_c + \Delta M$ and different values of $n_B/\rho_0$, where $\rho_0$ is the normal nuclear density. The subtracted term, $\varepsilon(n_B, n_Q^B = 0)$, only shifts the energy density to the origin when the total baryon density is zero. When there is no deviation of quark mass from $M_N/N_c$ one may see that the global minimum occurs at $n_Q^B = 0$ or, for high density, at $n_Q^B = n_B$, which means that there are only two phases: entirely quarks and entirely nucleons, separated by a first order phase transition. For the case where $\Delta M \neq 0$ the low density matter is nucleonic and at high density it is Quarkyonic, and they are separated by a first order phase transition. In this figure we have considered only vector interactions, therefore the energy density is given by the first two lines of Eq.~\ref{EnDens}, and the parameters are $M_N = 939$ MeV, $M_v = 783$ MeV and $g_v = 8$.

\section{Quarkyonic Matter in a Phe\-no\-me\-no\-lo\-gi\-cal Model of Quark In\-te\-rac\-tions}

While a description in terms of free constituent quarks may be reasonable at very high density, it is certainly not at the low densities at which the filled shell of quark degrees of freedom first appear.  Interactions are important.  The final model we consider will be to allow these quark degree of freedom to have a phenomenological relationship between density and chemical potential.  In particular,  at low density, we will assume that
 \begin{equation}
  \mu_Q = m_Q + {{n_B^Q} \over \Lambda^2}\,, 
  \label{mod_muQ}
 \end{equation}
 where ${n_B^Q}$ is the baryon number associated with quarks, and $\Lambda$ is parameter of order of $\Lambda_{\text{QCD}}$.  In such a model, we will assume that there is no mean field associated with the quarks.  Although this look like a harmless low density expansion, it makes the assumption that there is not free kinetic energy term associated with quarks which would be non-analytic for small $n_B$ and would contribute a term proportional to $(n_B^Q)^{1/3}$ at high density. Numerically this is implemented as 
 \begin{eqnarray}
  \mu_Q^B &=& N_c\left[\left(m_Q + \frac{a}{m_Q^2}n_B^Q\right)\Theta(r - n_B^Q) \right.\nonumber\\
  && \left.+ \left(\frac{3\pi^2}{2}\right)^{1/3}(n_B^Q)^{1/3}\Theta(n_B^Q - r) \right]\,,
 \end{eqnarray}
where $r$ is the solution of the equation $\left(m_Q + \frac{a}{m_Q^2}r\right)  -\left(\frac{3\pi^2}{2}\right)^{1/3}r^{1/3} = 0$.
The linear dependence of quark chemical potential with the density was the simplest form we could find that allowed for a continuous transition between nuclear matter and Quarkyonic matter, and such behavior is favored by the data extracted from neutron stars concerning the high density matter equation of state.
 
Since there is no free kinectic energy associated with the quarks, it is not possible to define directly a connection between the quark density $n_Q^B$ and the quark fermion momentum $k_Q$. We follow then the approach suggested in Ref.~\cite{Duarte:2021tsx}, by requiring the ghost potential $\mu_G$ is related to the quark chemical potential $\mu_Q$ as
\begin{equation}
  \mu_G =  \mu_G^* + g_vV^0\,, \;\;\;\;\;\;\;\;\;\;\; \mu_G^* = N_c \mu_Q
\end{equation}
where $V^0 = (g_v/M_v^2)n_v$ is the background vector potential, and the quark chemical potential is given by Eq.~\eqref{mod_muQ}.  
If one considers the possible larger mass scale of the constituent quarks due to the compressed mixture of the nucleon and resonances,  the ghost chemical potential should be shifted by the threshold mass: 
 \begin{equation}
  \mu_G = \mu_G^* + g_v V^0\,, \;\;\;\;\;\;\;\;\;\;\; \mu_G^* = N_c(\mu_Q - \Delta M),
 \end{equation}
so that we begin having ghost states when the quark states appear~\cite{Duarte:2021tsx}.
In this approach the ghost Fermi momentum in defined numerically as $k_G = \sqrt{(\mu_G^*)^2 - m_G^2}$, and the nucleon Fermi momentum, $k_N$, is defined in terms of the total baryon density relation $n_B = n_N - n_G + n_Q^B:$
\begin{equation}
 k_N = \left[\frac{3\pi^2}{2}\left(n_B + \frac{2}{3\pi^2}k_G^3 - n_Q^B\right)\right]^{1/3}.
\end{equation}

\begin{figure*}[!]
\begin{center}
\subfigure{\includegraphics[scale=0.4]{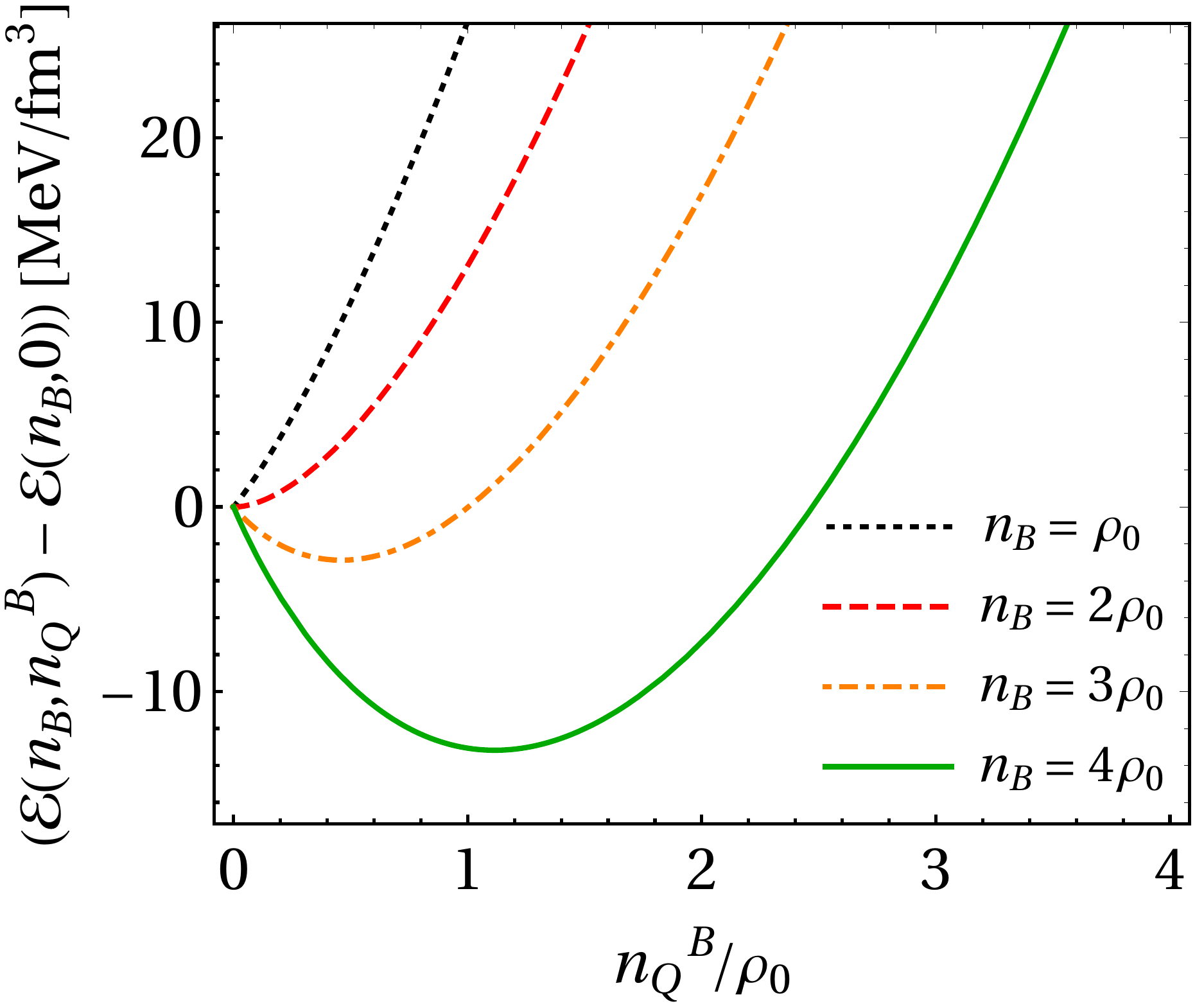}}\hspace{1cm}
\subfigure{\includegraphics[scale=0.41]{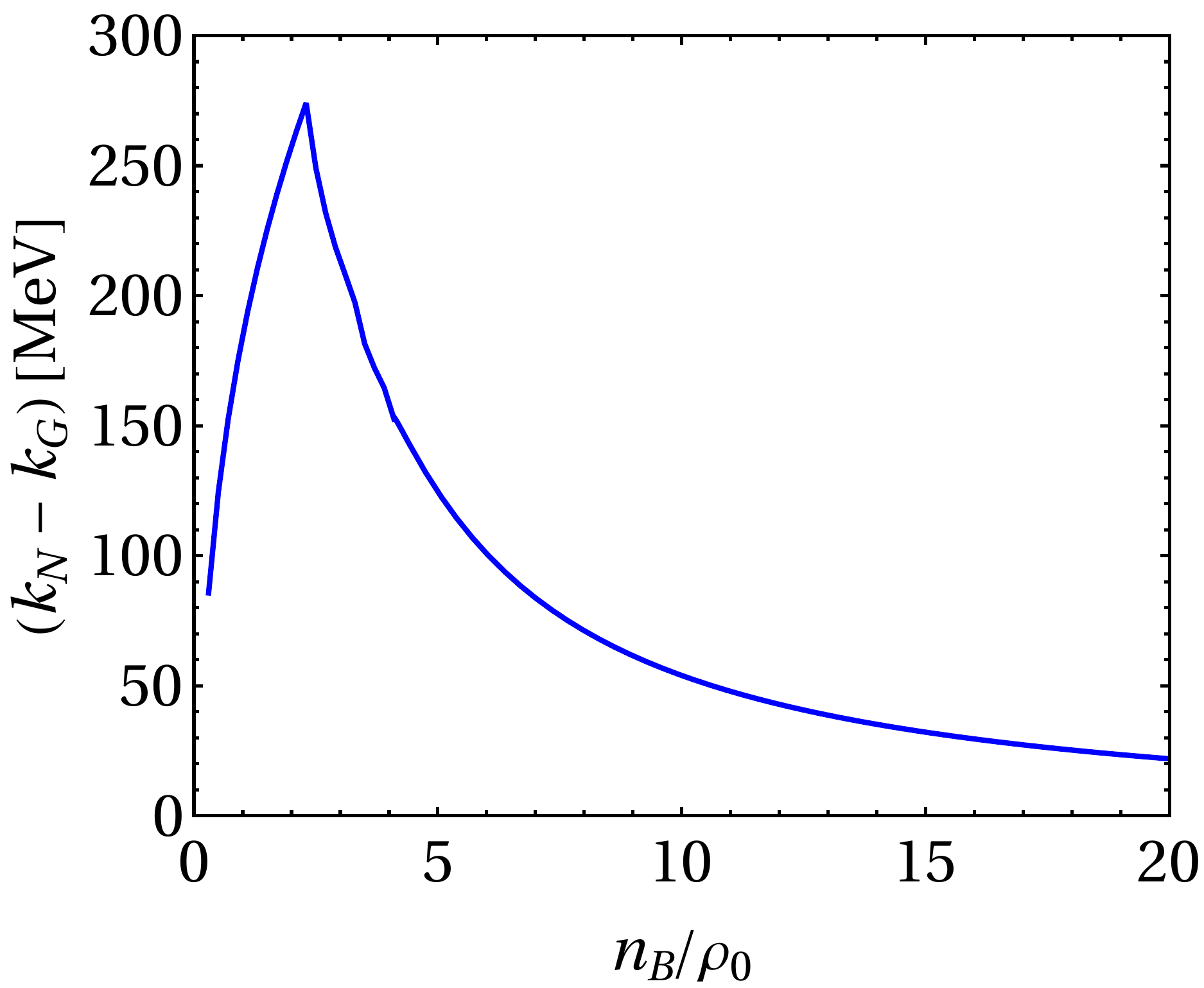}}
\caption{Left: Energy density as a function of quark density $n_Q^B$, for an interacting quark model and different values of total baryon density $n_B$, including only vector interactions in the nucleon sector.  Right: Shell thickness, defined as the difference between the nucleon and ghost Fermi momenta $k_N - k_G$.}
\label{fig2}
\end{center}
\end{figure*} 

A continuous transition from nuclear to Quarkyonic matter may be observed by choosing $\Lambda^2 = m_Q^2/a$, where $a$ is a parameter to be adjusted. 
By using $a = 1.2$, $\Delta M = 70$ MeV, and $g_v = 6$ one may observe from the left panel of Fig.~\ref{fig2} that between 2-3$\rho_0$ a global minimum appears in the energy density as a function of the quark density $n_Q^B$ close to the origin, which is a characteristic of smooth transitions. As the density increases and the quarks start to take the low phase space, the correspondent appearance of the ghosts guarantees that quarks inside nucleons do not occupy the same states where there are quarks, leading to a dynamical formation of a nucleon shell. This is illustrated on the right panel of Fig.~\ref{fig2}: for low densities, any increment of the total baryon density corresponds to an increase of the nucleon phase space, whose Fermi momentum is $k_N$. Around $n_B \sim 2.2\rho_0$ quarks are saturated, and the shell thickness can be defined as the difference between the nucleon and ghost Fermi momenta, $k_N - k_G$. As the density increases, the shell gets thin and it is expected to disappear at asymptotically large densities when the quarks eventually deconfine.

In Ref.~\cite{Koch:2022act} it was shown that in the isospin symmetric quasi-particle model of baryon-quark mixture, a con\-fi\-gu\-ra\-tion in which the Fermi sea is filled with confined baryons surrounded by a shell of deconfined quarks is energetically favorable compared to quarkyonic matter. This scenario, called {\it Baryquark matter}, represents another possible realization of the Pauli exclusion principle, where the momentum shell structure is also generated dynamically, being particularly preferred in transport simulations.

The mean-field approach considered in this work may generate Baryquark matter if one defines the ghost and the mean-field potential of the nucleons in different ways. The ghost field for the Baryquark matter should be defined to exclude the quarks occupying the inner part of the quark Fermi sea and the weak interaction should be assumed for the constituent quark sea. For the nucleon interaction side, one may have to consider the attractive scalar mean-field potential beyond linear density order to match the attractive potential considered in Baryquark matter~\cite{Koch:2022act}. Also, as the excluded volume approach in the baryon sector~\cite{Rischke:1991ke, Vovchenko:2015vxa} generates both Quarkyonic and Baryquark configuration~\cite{Jeong:2019lhv, Koch:2022act}, the strong repulsive vector channel will be required for the nucleon potential. The details of model construction will be pursued in the future work on the transport modelling.

\section{Summary and Conclusions}

In this work, we extend the idea of a Quarkyonic matter description in terms of a field theory composed of nucleons, ghosts, and quarks. The ghost field is introduced to prevent the overcounting of states where the quarks inside of nucleons might occupy the same physical states as those associated with the states in the filled quark sea.
To describe Quarkyonic matter, we reviewed the simplest ingredients in the mean field theory framework required by the consideration of ghost field and subsequent modification of the interaction between the nucleons and quarks.
In the simple phenomenological point of view, the filled quark states is approximated as the compressed state of the quasi-nucleons and its higher mass resonances, where the constituent quark wave functions start to interfere with each other. In this point of view, it is natural to take the nonzero threshold mass $\Delta M$ in the modelling of Quarkyonic matter. The physical role of the  threshold mass becomes clear in the analysis of the free quark sea case. If $\Delta M = 0$, there is a first order phase transition from nuclear to quark phase. If one considers $\Delta M\neq 0$ from the compressed matter states of the nucleons and resonances a kind of first order transition to Quarkyonic matter is obtained.

On the other hand, as the interactions of the first quark degrees of freedom appearing at low intermediate densities are significant, we implemented a relation between the quark chemical potential and the correspondent quark density in the form of a low density expansion, which eventually converges to the free gas approximation at high densities. Under this approximation, the Quarkyonic phase appears around $n_B \sim 2.2\rho_0$. As the baryon number density increases, the shell thickness gets thinner and the thin-shell limit is expected to appear as claimed in the large $N_c$ limit.  This assumption allows a continuous transition between nuclear and Quarkyonic matter that is suitable to take into account the rapid rise of the sound velocity required in the density evolution of the equations of state adequate to describe the observations of neutron stars. Although it is a model dependent result, the transition may occur at relatively low densities (around $n_B \simeq 2.0\rho_0$) which implies that the Quarkyonic phase can be explored not only in the context of the astrophysical observations but also in the future planned heavy-ion collision experiments \cite{Senger:2022bjo, Lovato:2022vgq}.

\begin{acknowledgments}
K.S. Jeong, and L.  McLerran was supported by  the U.S. DOE under Grant No. DE-FG02- 00ER4113.  
K.S. Jeong acknowledge the support from the Simons Foundation on the Multifarious minds Grant No. 557037 to the Institute for Nuclear Theory. 
D.C. Duarte also acknowledge Funda\c{c}\~ao de Amparo \`a Pesquisa do Estado de S\~ao Paulo (FAPESP) grant No. 17/26111-4.  
S. Hernandez-Ortiz was supported by the Department of Energy Nuclear Physics Quantum Horizons program through the Early Career Award DE-SC0021892.
Larry McLerran acknowledges useful insight provided by Jin-feng Liao, Rob Pisarski and Sanjay Reddy.
\end{acknowledgments}

\end{document}